\documentclass[12pt]{article}
\usepackage{cite}
\usepackage{amsmath,amssymb,amsfonts}
\usepackage{caption}
\usepackage{algorithmic}

\usepackage[margin=1in]{geometry}

\usepackage{authblk}

\usepackage{fancyhdr}

\fancypagestyle{AllPages}{
}
\fancypagestyle{FirstPage}{
\rfoot{doi: 10.1109/ACCESS.2018.2882777}
\lfoot{\copyright 2018 IEEE}
}

\usepackage{hyperref}

\usepackage{float}

\usepackage[pdftex]{graphicx}
\graphicspath{{fig/}}

\begin{document}

\title{An Emergent Space for Distributed Data with Hidden Internal Order through Manifold Learning}
\author[1, 2]{Felix P. Kemeth\thanks{These authors contributed equally to this work.}}
\author[1, 2]{Sindre W. Haugland$^*$}
\author[3, 4]{Felix Dietrich}
\author[5]{Tom Bertalan}
\author[1]{Kevin H\"ohlein}
\author[6]{Qianxiao Li}
\author[7]{Erik M. Bollt}
\author[8]{Ronen Talmon}
\author[1]{Katharina Krischer}
\author[2, 3, 4]{Ioannis G. Kevrekidis}
\affil[1]{Physik-Department, Nonequilibrium Chemical Physics, Technische Universit\"{a}t M\"{u}nchen,
  James-Franck-Str. 1, D-85748 Garching, Germany}
\affil[2]{Institute for Advanced Study - Technische Universit\"{a}t M\"{u}nchen,
  Lichtenbergstr. 2a, D-85748 Garching, Germany}
\affil[3]{The Department of Chemical and Biological
  Engineering - Princeton University, Princeton, NJ 08544, USA}
\affil[4]{Department of Chemical and Biomolecular Engineering, Department of Applied Mathematics and Statistics, Johns Hopkins University and JHMI} 
\affil[5]{Department of Mechanical Engineering - MIT, Cambridge, MA 02139, USA}
\affil[6]{Institute of High Performance Computing,  1 Fusionopolis Way, \#16-16 Connexis North, Singapore 138632, Singapore}
\affil[7]{Department of Mathematics, and Department of Electrical and Computer
  Engineering, Clarkson Center for Complex Systems Science, Clarkson University, Potsdam, NY 13699-5815, USA}
\affil[8]{Department of Electrical Engineering - Technion - Israel
  Institute of Technology, Technion City, Haifa, Israel 32000}

\date{\vspace{-5ex}}

\maketitle

\thispagestyle{FirstPage}



Data Mining, Diffusion Maps, Dimensionality Reduction, Nonlinear Dynamical Systems

\begin{abstract}
Manifold-learning techniques are routinely used in mining complex spatiotemporal data to extract useful, parsimonious data representations/parametrizations; these are, in turn, useful in nonlinear model identification tasks.
We focus here on the case of time series data that can ultimately be modelled as a spatially distributed system (e.g. a partial differential equation, PDE), but where we do not know the space in which this PDE should be formulated. Hence, even the spatial coordinates for the distributed system themselves need to be identified - to ``emerge from''- the data mining process. 
We will first validate this ``emergent space'' reconstruction for time series sampled {\em without space labels} in known PDEs; this brings up the issue of observability of physical space from temporal observation data, and the transition from spatially resolved to lumped (order-parameter-based) representations by tuning the scale of the data mining kernels.
We will then present actual emergent space ``discovery'' illustrations.
Our illustrative examples include chimera states (states of coexisting coherent and incoherent dynamics), and chaotic as well as quasiperiodic spatiotemporal dynamics, arising in partial differential equations and/or in heterogeneous networks. 
We also discuss how data-driven ``spatial'' coordinates can be extracted in ways invariant to the nature of the measuring instrument. Such gauge-invariant data mining can go beyond the fusion of heterogeneous observations of the same system, to the possible matching of apparently different systems. 
For an older version of this article, including other examples, see \href{https://arxiv.org/abs/1708.05406}{https://arxiv.org/abs/1708.05406}.

\end{abstract}

\lfoot{\copyright 2018 IEEE}



\section{Introduction}
\label{sec:introduction}

In 1979, when the American embassy was vacated in Tehran, sensitive documents were not incinerated; they
 were instead ``strip-shredded'' and considered destroyed. Yet local carpet-weavers painstakingly put them 
together again~\cite{AdrianChen2009}. 
In 2011 DARPA issued a ``shredder challenge'': reconstructing five shredded pages (a handwritten page,
 a picture etc.); the \$50,000 prize was claimed a few weeks later~\cite{darpa}.

The subject of this paper is the extraction of useful information {\it implicit} in apparently spatially disorganized data, such as the actual physical order of the paper shreds; this can be useful for winning the DARPA challenge, or, as we show below, 
for visualizing and hopefully better understanding spatiotemporal simulations.
Extracting useful information, such as the dimension and geometry of the underlying physical space, is crucial in identifying a consistent distributed dynamic model. 
Establishing the correct global sequence of temporal observation windows, as well as the number and nature of state variables in this consistent dynamic model is also crucial for the system identification task.
In this paper we use nonlinear data mining tools (in particular, versions of diffusion 
maps~\cite{COIFMAN20065, Coifman24052005, Nadler-2006, SINGER2008226, dsilva16_data_driven_reduc_class_multis})
to extract these types of information from spatially/temporally disorganized data by exploiting the intrinsic variabilities in the recorded data sequences. 
These extracted variabilities are the key to defining new spatial coordinates for 
the representation and modeling of the system.

In classical mechanics, generalized coordinates define a frame in which to represent the configuration of a 
system with respect to a reference state.
Many possible choices of generalized coordinates exist, and there is an ``anthropic'' motivation in choosing
 them so as to make the formulation, or the solution, of the equations of motion in these coordinates easier.
Data-driven coordinate discovery provides an even broader set of alternatives. 
Ideally, such a set of coordinates would be intrinsic to the phenomenon observed and independent of 
the particular nature of the measurement entity.

The focus in this paper is threefold: We will start with the observability and reconstruction of physical space from disorganized (spatially unlabelled) observations of spatially distributed processes.
The extent to which we can ``rediscover'' the missing physical space validates the approach.
In the case of networks, where no obvious physical space exists, we will demonstrate the identification of a ``surrogate'' embedding space, useful in reducing the network dynamics, 
demonstrate how it emerges from the network dynamic data, and discuss its possible physical interpretation in terms of network heterogeneity.
Motivated by this heterogeneous network example, we choose the term ``Emergent Space'' for such spatial embeddings.

Physical dynamics can be observed at several different scales.
Our second focus involves the adaptation of data mining to the degree of feature coarse-graining we are interested in: that is, to the scale of the observer.
Investigating chimera states (hybrid states of coherent and incoherent dynamics) for a partial integro-differential equation, 
we show how we can move from recovering space on a fine scale to the extraction of 
a more lumped description in terms of more coarse-grained order parameters.

Lastly, we focus on exploring the use of different observation modalities. 
Our illustrative example involves low-dimensional spatiotemporal dynamics,
namely modulated traveling waves (quasiperiodic solutions) of the Kuramoto-Sivashinsky equation, and we start by reconstructing the underlying toroidal attractor.
We then show how one can recover the same emergent space reconstruction from different types 
of (disorganized) measurements.
We draw connections between this and traditional dynamical system ``embedology''~\cite{sauer-1991}: the theorems by Whitney~\cite{eells-1992}, Nash~\cite{nash-1954}, and Takens~\cite{takens-1981}.
If two systems are thus found to be observations of each other, we illustrate, in a simple example, how spectral techniques (gauge-invariant data mining, but also Koopman-operator techniques~\cite{budisic-2012,bollt-2018b}) can guide us in matching the two systems.
Being able to consider (in principle) all possible diffeomorphisms of a given representation of a system, leads naturally to a discussion of which representation to choose to work with and report on. For which variables should we try to identify a dynamic model?
This, in turn, leads to some simple arguments about the interplay of data mining (and, more generally, machine learning) and systems modeling, opening up several research directions, some of which we outline.

Details on the individual models and algorithms are given in the Supplementary Information~\cite{supplement}.

\section{On Diffusion Maps}
In recent years, many nonlinear manifold learning techniques have
been proposed for embedding apparently high-dimensional nonlinear phenomena in lower-dimensional spaces.
Examples include Isomap~\cite{Tenenbaum2319}, locally linear embedding~\cite{Roweis2323}, 
Laplacian eigenmaps~\cite{Belkin}
and our method of choice here: Diffusion Maps~\cite{Coifman24052005,Singer22092009,COIFMAN20065,Coifman08}.

Assume a collection of $N$ real-valued time series segments $\{a_i\}$, $i = 1, \dots, N$.
Each sampled time series $a_i$ is a $T$-dimensional vector, containing the
value of one recorded variable at $T$ discrete points in time, with $T$ depending on the
sampling rate and the time window considered 
(longer vectors sampling more than one variable at each moment in time can be
similarly constructed).
Thus, each (time series) vector can be regarded as a point in a $T$-dimensional
Euclidean space, with all time series together forming a point cloud embedded in $T$ dimensions.
We assume that this point cloud lies close to a smooth manifold $\mathcal{M}$, embedded in $\mathbb{R}^T$.
Calculating all pairwise Euclidean distances between the points
yields, through a diffusion kernel,  
a symmetric $N\times N$ matrix, whose spectral properties 
can reveal intrinsic structures in this cloud.

A diffusion kernel weighs the Euclidean distances between
points that are close in $T$-dimensional space (i.e., between similar time series segments), 
much more strongly than those pairs of points at larger Euclidean distances.
This effectively embodies a random walk on the data, where the
probability of jumping from point $a_i$ to $a_{i'}$ is large if their distance
is small, and vice versa.
The diffusion kernel on $\mathbb{R}^T$ is defined as
\begin{equation*}
  k\left(a_i,a_{i'} \right) = e^{-\frac{\|a_i-a_{i'}\|^2}{2\epsilon}}\,.
\end{equation*}
Here, the scale parameter $\epsilon$ is used to tune
the rate of decay of the kernel compared to a characteristic scale present in the data.
For relatively small $\epsilon$, the kernel decays fast, and only pairs of points
very close in the Euclidean distance are significantly weighted.

Using the similarity measure given through the kernel, we construct a graph between all data points. For a given scale parameter $\epsilon>0$, the connectivity between two points $a_i,a_k\in \mathbb{R}^T$ is stored in the kernel matrix $K_{ik}=k(a_i,a_k)$. The Diffusion Maps concept is based on the convergence of the normalized graph Laplacian on this graph to the Laplace--Beltrami operator on the manifold $\mathcal{M}$, in the limit $n\to\infty, \epsilon\to 0$. If the data points are not sampled uniformly on the manifold, the matrix $K$ has to be normalized by an estimation of the density,
\begin{eqnarray}
P_{ii}&=&\sum_{k=1}^n K_{ik},\\
\tilde{K}&=&P^{-\alpha} K P^{-\alpha}.
\end{eqnarray}

Here, $\alpha=0$ (no normalization \cite{Belkin}) can be used in the case of uniform sampling, and $\alpha=1$ otherwise \cite{COIFMAN20065}. The kernel matrix  $\tilde{K}$ then has to be normalized again, by the diagonal matrix $D_{ii}=\sum_{k=1}^n \tilde{K}_{ik}$, yielding the Markov matrix $A=D^{-1} \tilde{K}$. A non-linear parametrization (embedding) of the manifold is then given by a certain number $l$ of eigenvectors of $A$, scaled by their respective eigenvalue. The new embedding dimension $l$ can be much smaller than the previous ambient space dimension $T$, in which case the algorithm achieves dimensionality reduction.
Selecting the eigenvectors is not as straight-forward as for linear dimensionality reduction methods (Principal Component Analysis), but can be achieved, for example, by sorting the eigenvectors based on the absolute value of their associated eigenvalue and removing eigenvectors that are functions of previous ones~\cite{dsilva-2018}.

Changing the length scale $\epsilon$ can lead to very different representations
of the same data, as described in more detail in the section ``The scale of the observer'' in the paper.
For convenience, we use $\epsilon = \text{const} \cdot D_{\text{max}}^2$, with $D_{\text{max}}$ 
being the maximal distance in the data set.

While most of our results are obtained using a simple Euclidean-distance-based diffusion kernel, many extensions exist, such as vector diffusion
maps and (non)orientable Diffusion Maps~\cite{CPA:CPA21395};
we will also use \emph{anisotropic} diffusion kernels, 
based on the more refined, so-called Mahalanobis-like distance~\cite{SINGER2008226} (Section~\ref{sec:mtw}).
Recent Diffusion Maps research has progressed along a broad range of topics and
applications, 
including intrinsic modeling~\cite{talmon13_empir_intrin_geomet_nonlin_model}, 
reduction in multiscale dynamical systems~\cite{dsilva16_data_driven_reduc_class_multis}, 
multimodal data analysis~\cite{lederman15_learn_geomet_common_laten_variab}, 
and data organization~\cite{coifman11_harmon_analy_digit_data_bases},
just to name a few.

Diffusion maps help construct a nonlinear, data-based change of coordinates, embedding 
an intrinsically $d$-dimensional manifold in a low-dimensional Euclidean space. 
The extrinsic dimension of the manifold can be much larger than $d$, making this a suitable choice 
for dimensionality reduction.
In this paper, we use Diffusion Maps to discover intrinsic order contained in a data set, focusing on the relation between this order and the spatiotemporal dynamic modeling of the data.  
%

\begin{figure}[ht!]
  \centering
    \includegraphics[width=\columnwidth]{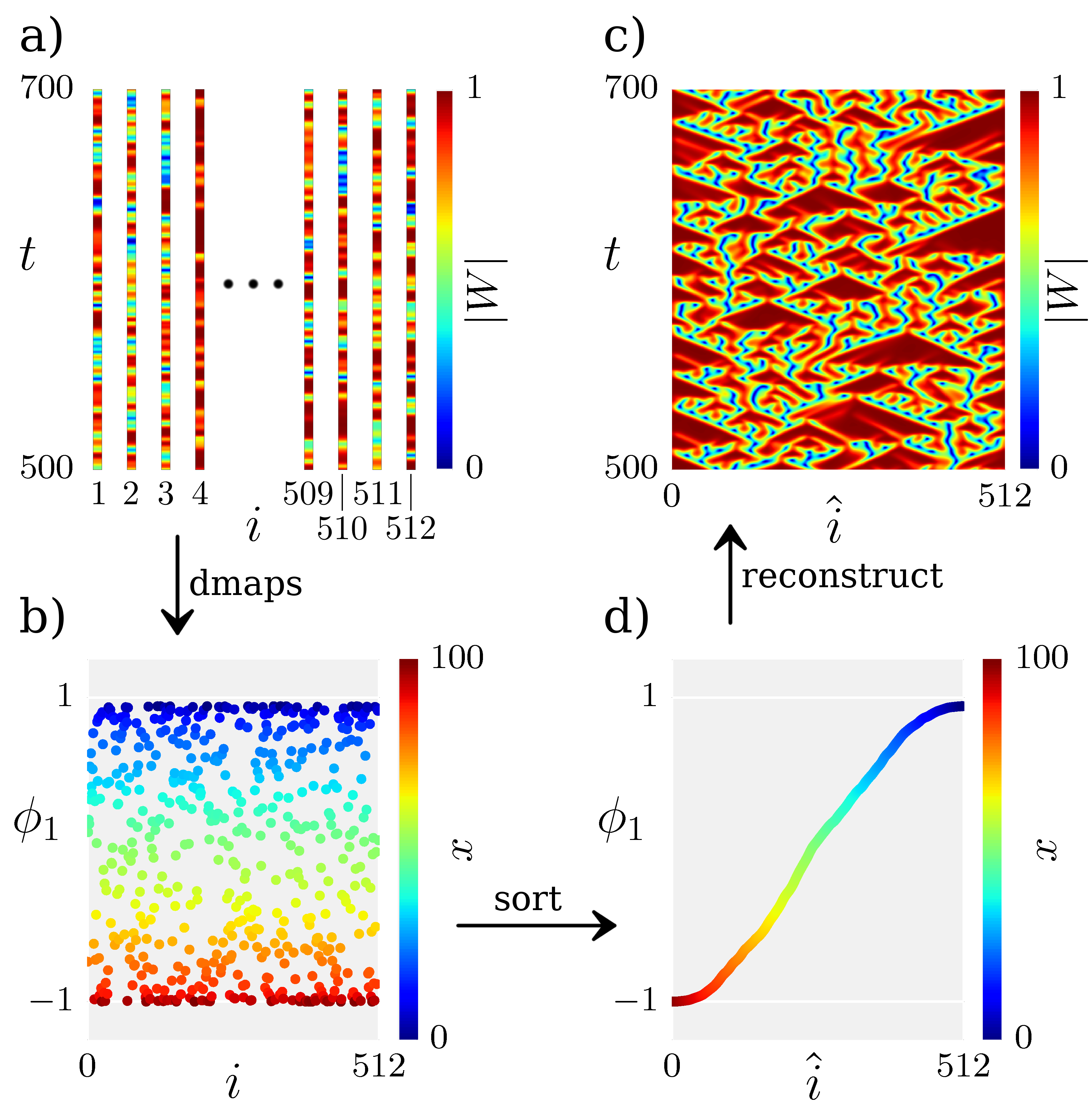}
    \caption{(a) Randomly shuffled time series segments (indexed by $i$) observed at different spatial locations in a simulation of the complex Ginzburg-Landau equation.
      The color corresponds to
      the modulus of the complex amplitude $W$. (b) First
      independent Diffusion Map coordinate obtained from the shuffled time
      series segments in (a).  (d) Sorting the time series in (a)
      according to their respective first Diffusion Map coordinate
      yields a new indexing, $\hat{i}$. 
      (c) The time series in (a) sorted according to this new, data-driven indexing $\hat{i}$;
      this identifies the correct topology of the original physical space, and reconstructs
      the original spatiotemporal simulation results.}
  \label{fig:1}
\end{figure}

\section{Recovering space from spatiotemporal data}
\label{sec:recov-space-from}

\begin{figure*}[!ht]
  \centering
    \includegraphics[width=0.98\textwidth]{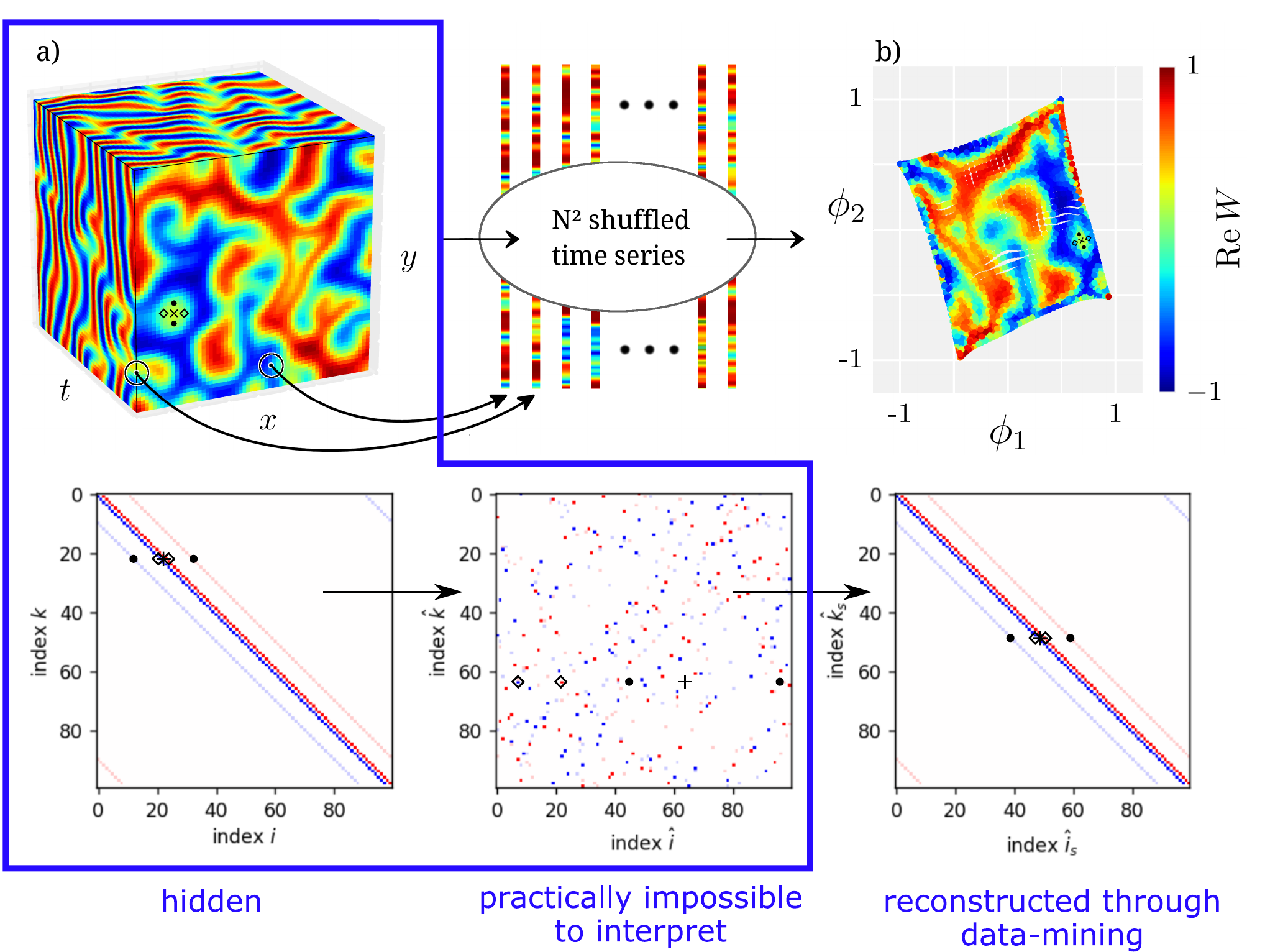}
  \caption{ Top row: (a) Spatiotemporal data obtained from numerical simulations 
of the complex Ginzburg-Landau equation in two spatial dimensions. 
The data is colored by the real part of the complex amplitude $W$. 
To demonstrate that we can reconstruct (a version of) the original physical space 
from disorganized, spatially unlabelled time series data, we randomly shuffle the $N^2$ time series segments before using Diffusion Maps.
  (b) The first two independent Diffusion Map coordinates, $\phi_1$ and $\phi_2$, resulting from data mining
  and colored as in (a); the ``rediscovered'' space is visually homeomorphic to the hidden ``true'' one. 
  Apart from a slight rotation, $\phi_1$ corresponds roughly to $-x$ and $\phi_2$ to $y$.
  Bottom row (schematic): in a finite difference discretization of the PDE, the ``hidden'' Jacobian would have a banded structure in the usual
  row-by-row numbering of the discretization points (a typical point five-point stencil is highlighted, left). This structure is lost upon random point labelling (middle), but recovered when the points are correctly reordered after data mining (right). 
  }
  \label{fig:2}
\end{figure*}
We start by investigating observations of spatiotemporal
chaos in one spatial dimension. In particular, we 
consider intermittency in the complex Ginzburg-Landau equation, a
nonlinear reaction-diffusion-type partial differential
equation for a complex variable $W(x,t)$, arising in the modeling of oscillatory systems~\cite{SHRAIMAN1992241,CHATE1996348,morales2012}. 
The complex Ginzburg-Landau equation (CGLE) in a re-scaled form reads
\begin{equation}
  \label{eq:cgle}
  \partial_t W = W + \left(1+ i c_1\right) \nabla^2 W - \left(1 + i c_2 \right) \left| W \right|^2 W,
\end{equation}
with real parameters $c_1$ and $c_2$.
The parameter values leading to our
spatiotemporal intermittency~\cite{SHRAIMAN1992241} are $c_1 = 0$ and $c_2 = -3$ (see the supplemental information for details on the integration methods used).

The repeated appearance 
of temporally synchronous spatial patches at seemingly random locations in space and time
is characteristic of this dynamical regime.
Following their initial emergence, these patches shrink in size
due to diffusion, giving rise to the triangular patterns 
in Fig.~\ref{fig:1}(c). 
The incoherent nature of the dynamics suggests that the (here $512$)
time series recorded at each individual discretization point,
each consisting of $1000$ steps,
are mutually different (there is no spatial periodicity)
Nevertheless, due to the smoothness induced by diffusion in the system, time series at points located 
close to each other in physical space tend to be similar: The Euclidean distance
between observations
at nearby points will be smaller than that between observations at points
far apart in physical space. This fact will be exploited in the Diffusion Map process.
In Fig.~\ref{fig:1}(d), it can be seen that the 
first Diffusion Map coordinate $\phi_1$ associates (is one-to-one) with the spatial coordinate~$x$: 
Each data point (each time series) has its own entry in the $\phi_1$ vector; here large entries
correspond to time series observed at 
small $x$, while small entries correspond to observations at large $x$ (the original physical space was discovered ``flipped''!).
That $\phi_1$ (which is one-to-one with $x$) turns out to be by far the most dominant mode, 
suggests that the data mainly vary along a single ``distributed'' direction; 
alternatively, that the
system can be described using one spatial dimension. 
Note that for the kernel scale $\epsilon= 1.0 \approx 5.7\cdot 10^{-4} D_{\text{max}}^2$ 
chosen, with $D_{\text{max}}$ being the maximal Euclidean distance between our time series,
only a few nearest neighbors are effectively considered in the
Diffusion Map computation. (The effects of the kernel scale are treated in more 
detail in section 
``The scale of the observer: Tuning the kernel parameter''.)

Shuffling the individual time series, c.f. 
Fig.~\ref{fig:1}(a), i.e. shuffling the indices in the Diffusion Map matrix,
does not affect its eigencomputations.
The first and only significant Diffusion Map coordinate now appears
nonmonotonic with the (shuffled) index $i$, as shown in Fig.~\ref{fig:1}(b). 
However, simply sorting the entries of the Diffusion Map coordinate in increasing 
order (and changing the indices of the time series accordingly to the new index $\hat{i}$) recovers the 
original spatial arrangement, see 
Fig.~\ref{fig:1}(c). We thus argue that the dimensionality (one) and
the topology (the correct ordering of the points) of the physical space $x$ 
is contained in (is observable through and can be recovered from) the dynamic simulation data themselves, without spatial labels.

Shredding Fig.~\ref{fig:1}(a) vertically gave $512$ time series; shredding it in the
horizontal direction gives $1000$ spatial snapshots. They are mutually different,
yet the smoothness inherent in the time evolution implies that nearby time instances
will yield similar snapshots. 
The above procedure applies again: We can now recover from temporally shuffled snapshot
data the correct temporal sequence~\cite{dsilva-2015}.

Clearly, this approach to recovering physical space is not restricted to 
line segments. It is also applicable to one-dimensional systems
with periodic boundaries (rings, c.f. Fig. \ref{fig:4}), as well as to systems with two or more spatial dimensions. 
Indeed, our procedure so far can be thought of as the putting-back together of effectively one-dimensional  puzzles -- the puzzle pieces are very long in the vertical (time series) or in the horizontal (snapshot) direction. It is easy to rationalize that by comparing edges of space-time
patches (two-dimensional puzzle tiles) two-dimensional puzzles can also be reassembled.
For spatiotemporal intermittency in the complex Ginzburg-Landau 
equation with one spatial dimension and periodic boundary conditions, 
one finds that the first two Diffusion Map coordinates define a circle.
Shuffling the data indices, applying Diffusion Maps and sorting the shuffled data along 
that circle, one recovers the original ring ordering modulo a rotation and possibly
a reflection (not shown).

Fig.~\ref{fig:2}(a) illustrates the recovery of two-dimensional physical space from dynamical 
(time series segment) data.
It depicts
spatiotemporal intermittency on a square, with zero-flux boundary conditions, again for the complex Ginzburg-Landau equation (eq.~\ref{eq:cgle}).
Recording individual time series at each point of a 64 by 64 discretization mesh, 
and applying Diffusion Maps to these 4096 data segments, one finds two independent 
Diffusion Map coordinates $\phi_1$  and $\phi_2$.
Embedding the data points in $\left(\phi_1,\phi_2\right)$ space, and coloring them by the real part of the 
complex amplitude $W$ at $t=0$ shows that the two Diffusion Map coordinates indeed
span a space that is one-to-one with the original physical space --
by visual inspection, apparently a homeomorphism, see Fig.~\ref{fig:2}(b).

It is smoothness (similarity of recorded time series from nearby
points), that lets diffusion 
maps discover the right two-dimensional parametrization of the data and yield their
actual relative positions.
Neither the dimension of physical space, nor the physical location of each 
discretization (and thus, observation) point were used -- everything
is contained in (observable through, recoverable from) just the recorded dynamic data.


\section{Emergent Space Reconstruction in Network Dynamics}
\label{sec:recov-param-space}

Nodes in a network will, in general, not correspond to points in a low-dimensional
physical space (all-to-all coupled networks are an obvious example).
Our approach can help discover an effective embedding space, based on the 
dimensionality and the intrinsic geometry of the node \emph{dynamics}.

Consider a neuronal network that arises in modeling the pre-B\"{o}tzinger complex~\cite{Rubin2002,Laing2012,Choi2016}. 
The states of these Hodgkin-Huxley-type neurons (a membrane potential $V$ and a channel variable $h$) oscillate periodically.
As in previous studies~\cite{Choi2016}, the neurons are heterogeneous: Each is characterized by a different value of the intrinsic kinetic parameter  $I_{app}^i$.
\begin{figure*}[ht!]
  \centering
   \includegraphics[width=0.48\textwidth]{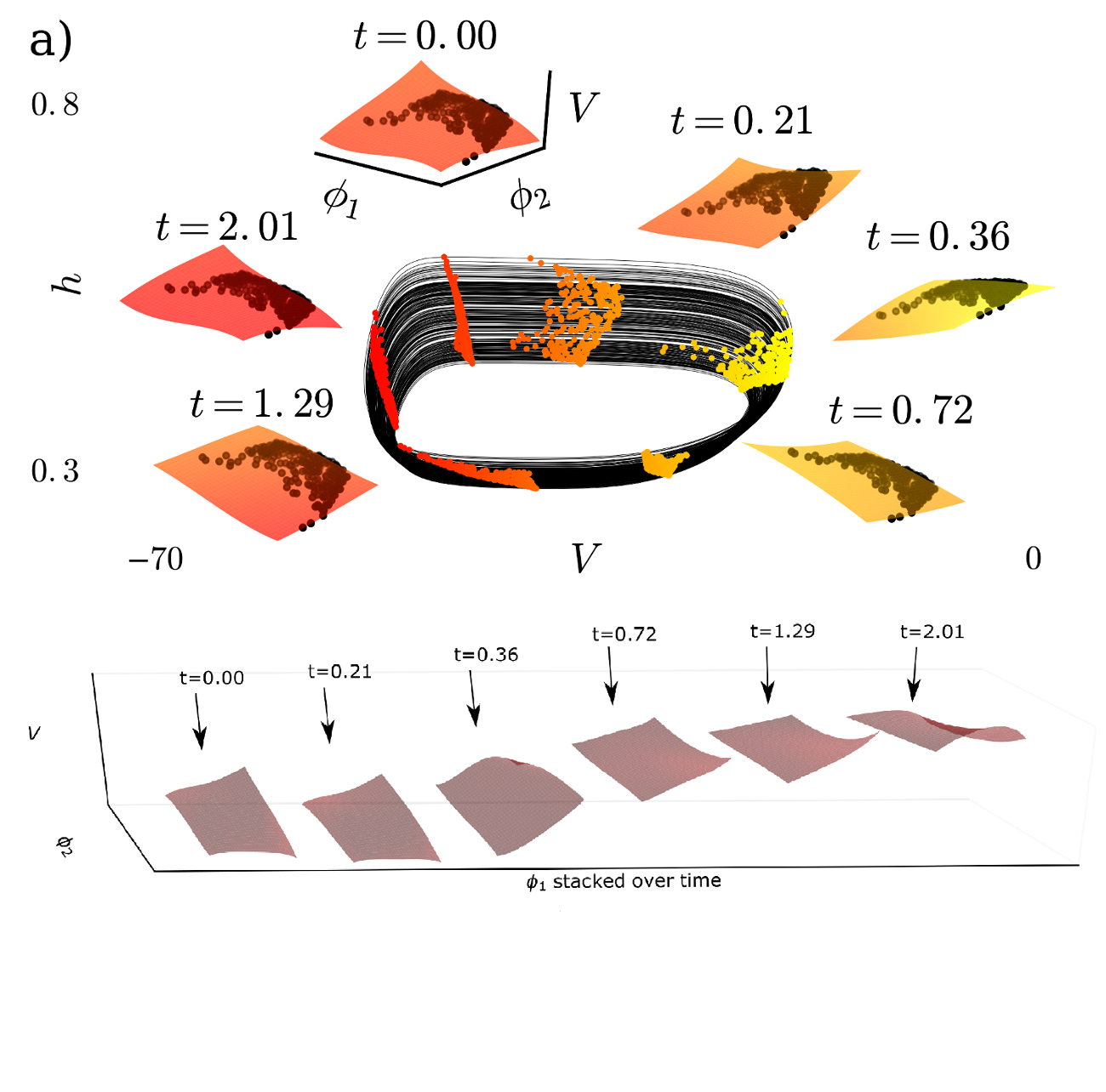}
    \includegraphics[width=0.48\textwidth]{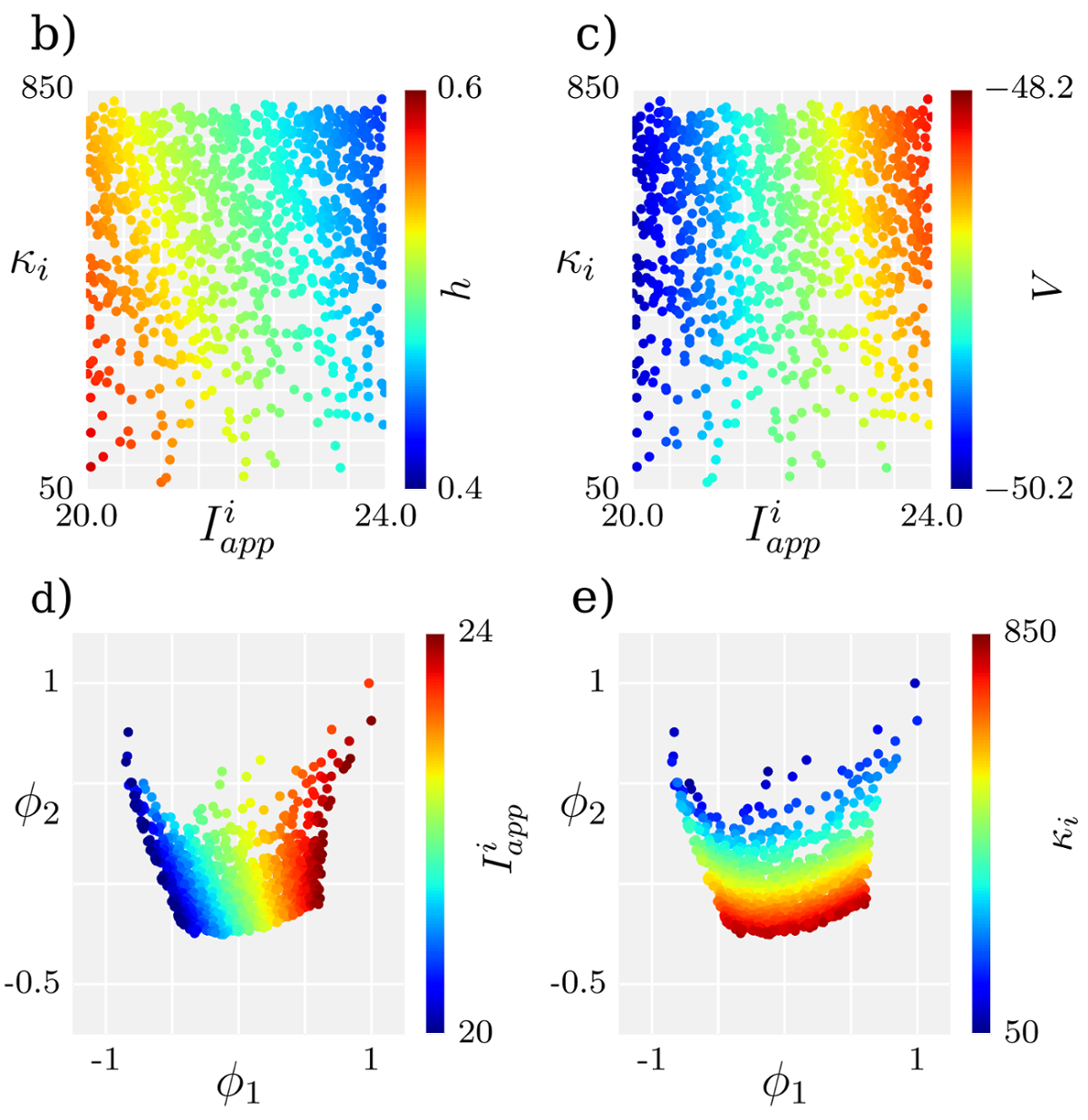}
  \caption{(a) Top: The closed curves in the center show temporal evolution of the two variables $h$ and $V$ of
    the 1024 individual oscillators in the Chung-Lu type network of 
    pre-B\"{o}tzinger neurons. For six representative temporal snapshots along the trajectory
    (indicated by the color of the point clouds on the attractor)
    the potential $V$ is  plotted as a function of the first two Diffusion Map coordinates $\phi_1$ and
    $\phi_2$. The value of $V$ for each neuron is marked by a dot. 
    The apparent smooth dependence of $V$ on the two Diffusion Map coordinates (~\cite{Choi2016}) strongly suggests that 
    the network could be modelled as an irregular grid discretization of a PDE in $\phi_1$ and
    $\phi_2$ (whose time evolution is sketched around as well as below the attractor).
    (b,c) Each oscillator as a function of the two heterogeneous parameters
    $\kappa_i$ and $I_{app}^i$, colored with the variable $h$ and the voltages $V$ at $t=0$. 
    (d,e) The two independent Diffusion Map coordinates $\phi_1$ and $\phi_2$, 
    colored by applied heterogeneous current $I_{app}^i$ and $\kappa_i$.} 
  \label{fig:3}
\end{figure*}

The neurons are also heterogeneous through their connectivity: They are not all-to-all coupled, but form a Chung-Lu-type network~\cite{Laing2012_2, supplement}. 
The number of connections to a neuron $i$, its degree $\kappa_i$, varies strongly across the neurons,
making it thus a second, \emph{structural}, heterogeneity. 
The (synchronized) temporal evolution of our network of $1024$ neurons is depicted in 
Fig.~\ref{fig:3}(a). 
The neurons oscillate clustered in the $(V,h)$
plane, but with somewhat different phases and amplitudes each. Due to the heterogeneities,
their instantaneous values differ. 
Nevertheless, the dynamics of the ensemble of neurons can be 
well approximated by a smooth function of the two heterogeneous parameters 
$I_{app}^i$ and $\kappa_i$~\cite{Choi2016}. 
This is indicated by the color code in Figs.~\ref{fig:3}(b-c). 

\emph{Without prior knowledge of these heterogeneity parameters}, the two-dimensional
nature of the collective dynamics can be recovered from temporal 
observations only, using Diffusion Maps.
Using pairwise distances between node time series segments
yields two Diffusion Map coordinates that parametrize a two-dimensional ``variability manifold''. 
The kernel width was chosen as $\epsilon=e^{10}\approx 0.39 D_{\text{max}}^2$.
Embedding the nodes in the resulting two-dimensional diffusion space,
and coloring them by the heterogeneous 
parameter $I_{app}^i$, one observes an approximate one-to-one correspondence 
between this parameter and the first Diffusion Map coordinate (Fig.~\ref{fig:3}(d)).
The second direction, transverse to the first, correlates with 
the degree $\kappa_i$, see Fig.~\ref{fig:3}(e). 

The leading few eigenvectors following $\phi_2$ are harmonics of the first two (not shown),
indicating that only two major directions parametrize the 
variability of the node dynamics. 
Comparing these results with the outcome from the
section ``Recovering space from spatiotemporal data'', an analogy arises between the two 
heterogeneity parameters in the neuronal network, 
and the spatial axes recovered in the 2-D reaction-diffusion system above.
Data mining enables us, by extracting the dominant variabilities in the dynamics in both cases, to find our ``Emergent Space''
(an effective embedding space).
In the reaction-diffusion example, this Emergent Space is one-to-one with actual physical space; in the network problem, it \textit{plays the role} of a physical space (even though the latter does not really exist). 
Indeed, representing the behavior in terms of well-chosen basis function sets in these intrinsic variability
dimensions makes the network description analogous to that of a discretized PDE in the to leading Diffusion Map coordinates (and time).
This can be used to dramatically reduce the network model: a few collocation
polynomials in variability space, instead of tens of thousands of coupled ODEs
accounting for each node separately~\cite{Choi2016}.

\section{The scale of the observer: Tuning the kernel parameter}
\label{sec:tuning-scale-param}

\begin{figure*}[!ht]
  \centering
  \includegraphics[width=\textwidth]{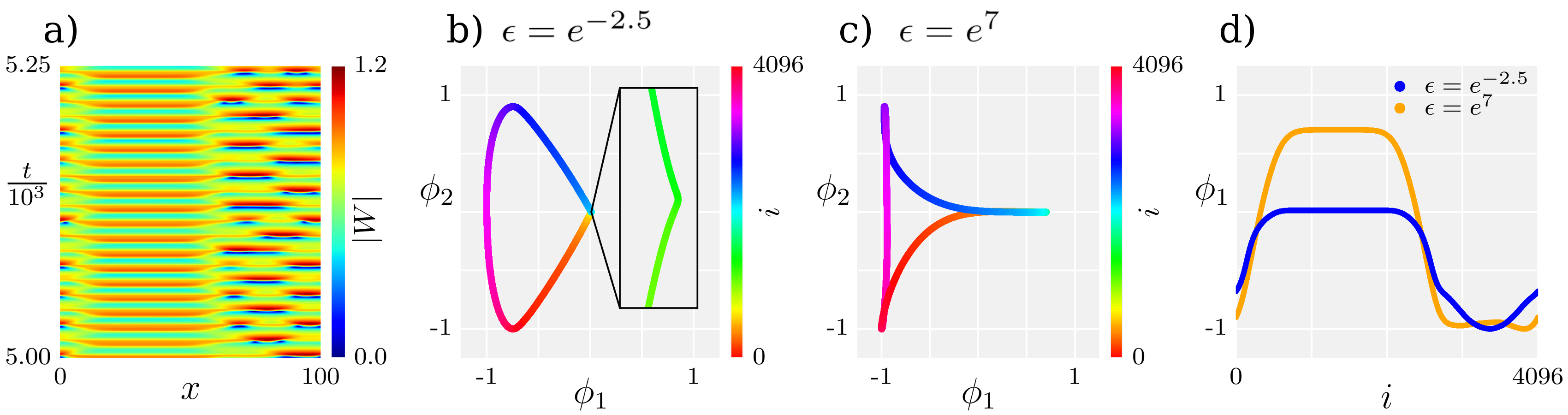}
  \caption{(a) Temporal evolution of a chimera state in a globally
    coupled version of the complex Ginzburg-Landau
    equation with one spatial dimension $x$ and periodic boundary 
    conditions. The pseudo-color corresponds to the modulus of the complex
    amplitude $W$. (b) The first two independent Diffusion Map coordinates 
    $\phi_1$ and $\phi_2$ for $\epsilon=e^{-2.5}\approx 3.5\cdot 10^{-5}D_{\text{max}}^2$, colored by the position $i$ along the spatial 
    coordinate $x$. (c) The first two independent Diffusion Map coordinates 
    $\phi_1$ and $\phi_2$ for $\epsilon=e^{7}\approx 0.46D_{\text{max}}^2$, colored by the position $i$ along the spatial 
    coordinate $x$.
    (d) First
    independent Diffusion Map coordinate $\phi_1$ for $\epsilon=e^{-2.5}\approx
    3.5\cdot 10^{-5}D_{\text{max}}^2$ and $\epsilon=e^7\approx
    0.46D_{\text{max}}^2$. (See text for discussion)}
  \label{fig:4}
\end{figure*}
In the previous examples, a specific choice of the kernel scale $\epsilon$ was made.
When recovering the Emergent Space for the PDE, we had to choose this $\epsilon$ to be very small, 
in the sense that only the nearest neighbors (only very similar time series) contribute to the computation. 
For the network, a coarser observation (a larger $\epsilon$, taking into account more than just nearest 
neighbors) was required to extract the two dominant variabilities.
In this section, we vary $\epsilon$ (the decay rate of the kernel, i.e. the scale of the observer) 
in order to explore how different features of our data are seen by observers
with different perception sensitivity. 
Increasing $\epsilon$ decreases the ability to discriminate between nearby time series, 
thus ``coarsening'' the observation.

As a model example, we observe a chimera state, 
that is, a dynamical hybrid state of coexisting coherence and
incoherence~\cite{kuramoto2002,abrams2004,panaggio2015,kemeth2016}. 
An example of such a state arises in a globally coupled version 
of the complex Ginzburg-Landau equation. 
A simulation in one spatial dimension is depicted in Fig.~\ref{fig:4}(a). 
Note that here we use periodic boundary conditions, so that the spatial axis
is in fact a ring.
This chimera state has an underlying two-cluster state: One of the two clusters develops 
incoherent dynamics while the other remains largely synchronized~\cite{Schmidt20152}.

By choosing the kernel scale $\epsilon=e^{-2.5}\approx 3.5\cdot 10^{-5}D_{\text{max}}^2$ -- very small, 
in the sense that only nearest neighbors, with very similar time series, contribute to the computation -- 
we are able to reconstruct the full circular spatial arrangement, as depicted in  Fig.~\ref{fig:4}(b). 
Due to the periodicity in $x$, two Diffusion Map coordinates are needed to embed the data. 
Note that this maps the coherent oscillations (by nature very similar) onto a dense cluster in Diffusion Map 
space. 
Nevertheless, by zooming in on this cluster, we find that the two Diffusion Map coordinates are still able to 
differentiate between the synchronous series, see the inset of Fig~\ref{fig:4}(b). 
This is possible since diffusion preserves a slight variation across the coherent cluster and therefore 
allows for distinguishing between the coherent time series.

We now vary $\epsilon$; the embedding in the first two Diffusion Maps eigenvectors 
for $\epsilon=e^7\approx 0.46 D_{\text{max}}^2$ is shown in Fig~\ref{fig:4}(c). 
This figure clearly shows that the observability of physical space is now lost
when observing the data in terms of its two most dominant variabilities.
There are three ``cusp like'' regions in the data; a finer-scale plot would
reveal intersections, which suggests that points at different physical locations (and
known to have different dynamics) appear the same at this observation scale.
Indeed, the entire coherent region effectively maps now to a point
(the rightmost point on this plot); and while the incoherent region
can be largely discriminated in terms of $\phi_2$, practically all its
constituents have the same $\phi_1$ value.
Observation in terms of only $\phi_1$ now reveals two plateaus (see Fig.~\ref{fig:4}(d));
the structure within the incoherent plateau observed at small $\epsilon$ is lost at this 
coarser observation level (large $\epsilon$).
Thus the system naturally coarse-grains from a one-dimensional PDE to a ``bistable'' system,
where two distinct scalar variables, two ``oscillator densities'' (the extent of the coherent and the incoherent regions)
interact. A small percentage of phase space is taken up by the transition regions (the fronts)
between the two plateaus. 
In a sufficiently large spatial simulation one might expect the extent (and the importance) of these transitions to 
be practically negligible.
This perception of clustering is consistent with earlier studies, which have shown that the 
dynamics of the type-II chimera state can be well approximated by a modulated-amplitude 
two-cluster state~\cite{Schmidt2015,Schmidt20152} - not a PDE anymore, but two ODEs.

Note that, as in the example of the pre-B\"{o}tzinger neurons, the
Euclidean distances between time series in the same $\phi_1$ cluster
can be quite large; yet observing these different time series through
the right observable (their $\phi_1$ component) helps cluster them meaningfully.

For very small kernel scales, the graph connecting data points is practically disconnected.
Every data point is a ``dimension'' by itself (eigenvectors approximate indicator functions, with local support and eigenvalue one). As $\epsilon$ is increased, these many distinct dimensions start to interact and gradually merge
into a coarser, one-dimensional ``emergent space'': the ring, described by the two Diffusion Map coordinates 
that we see dominating for $\epsilon = e^{-2.5}$.

This description is one-to-one with the physical space of the simulation, which thus 
emerges as a natural descriptor of the data for a range of $\epsilon$ values (Fig.~\ref{fig:4}(b)).
Further coarsening destroys the observability of physical space from the data (as indicated by multiple loops in Fig.~\ref{fig:4}(c)), leading eventually to
the ``two-cluster'' description through $\phi_1$ only.
Interestingly, this showcases the transition (in the eye of the beholder) between
three distinct regimes, clearly characterized by different scales and different physical interpretations: the set of $4096$ distinct coupled oscillators, the one-dimensional PDE with periodic boundary conditions, and the two-cluster amplitude description.

\section{Observations of an attractor}
\label{sec:mtw}
In the previous examples, time series observations were used in order to discover parametrizations of 
variabilities intrinsic to the dynamics, yielding coordinates in which to embed the data. 
Yet, all our observations came from a single time window; 
we did not observe long-term variabilities {\em along the time direction}, important in the study of dynamics.

\begin{figure*}[!ht]
  \centering
  \includegraphics[width=\textwidth]{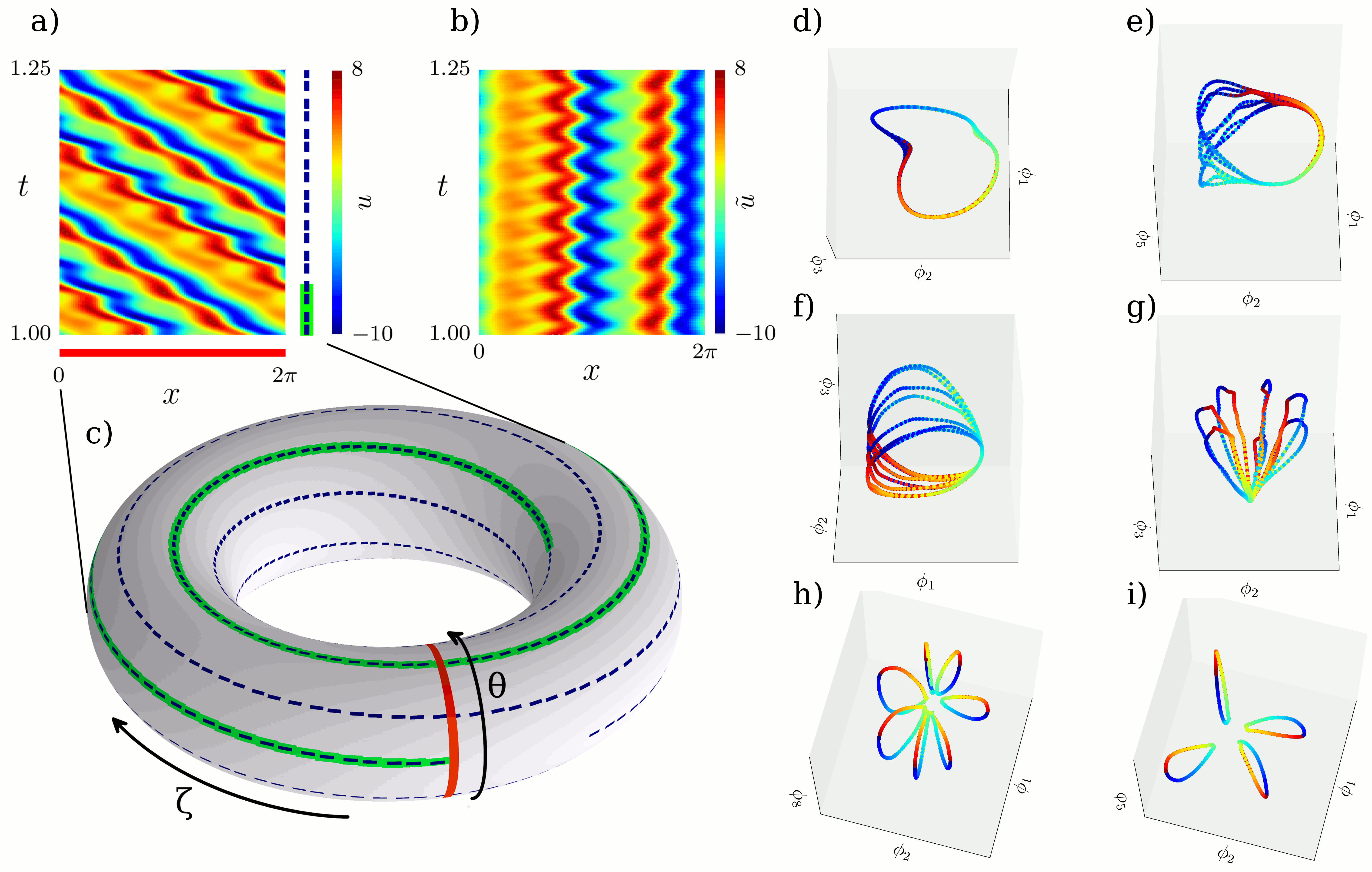}
       \caption{\label{fig:6}(a) Temporal evolution of
     $u$ in the 1-D Kuramoto-Sivashinsky
  equation with periodic boundaries, showing
  modulated traveling waves. (b) The dynamics of (a) in a co-rotating frame. The color
  corresponds to $u$.
  (c) Torus spanned by the phases with respect to traveling wave,
  $\theta$, and modulation, $\zeta$, respectively.
  Going along space (red line) only changes the former, whereas moving
  in time (green line, dashed blue line) changes both phases.
  (d) First three independent Diffusion Map coordinates obtained from overlapping time
  segments created with a time shift of $n=8$ and for $\epsilon =
  e^{-0.5}\approx 1.7 \cdot 10^{-2} D_{\text{max}}^2$, colored by spatial position.
  (e-h) Equivalents to (d) for
  $\epsilon \approx 3.8 \cdot 10^{-3} D_{\text{max}}^2$,
  $\epsilon \approx 2.1 \cdot 10^{-3}D_{\text{max}}^2$,
  $\epsilon \approx 2.0 \cdot 10^{-3}D_{\text{max}}^2$ and
  $\epsilon \approx 1.2 \cdot 10^{-3} D_{\text{max}}^2$,   
  respectively.
  (i) First three independent Diffusion Map coordinates obtained for $n=14$ 
  and $\epsilon = e^{-2.3}$, colored by spatial position.
  See discussion and interpretation of the images in the text.}
\end{figure*}

In this section we start using as data points partially overlapping time-series windows,
obtained during long-term simulations, in order to explore such temporal variabilities.
These window overlaps are chosen so that distances in the time- and the space-directions 
are (loosely) comparable.
Our illustrative example comes from observations of a modulated traveling wave (a quasiperiodic attractor) 
of the one-dimensional Kuramoto-Sivashinsky equation with periodic boundary conditions, shown in 
Fig.~\ref{fig:6}(a).

The Kuramoto-Sivashinsky equation is a fourth-order partial differential equation used to model spatiotemporal instabilities in a number of physical settings~\cite{HYMAN1986113,kevrekidis-1990}.
In this case, the dynamics possess two frequencies: one determined by the speed
of traveling along the periodic domain, and one coming from the temporal modulation.
The nature of the latter becomes apparent when observing the dynamics in a frame that is co-traveling 
with the wave, as depicted in Fig.~\ref{fig:6}(b).
Here, every spatial point oscillates differently, but with the same constant (modulation) frequency. 
Altogether, the dynamics (the short time window observations) at any particular point $x$ belongs to a 2-parameter family of possible behaviors, parametrized by the phases with respect to the traveling and to the modulation frequency.
In other words, the dynamics of the system lives on a (two-)torus, as schematically depicted in Fig.~\ref{fig:6}(c). 
There, the toroidal angle $\zeta$ denotes the phase with respect to the modulation, and the poloidal 
angle $\theta$ the phase with respect to the traveling of the wave.
%
A time-series window can be depicted as a short line-segment lying along this torus; all 
possible time segments constitute a two-parameter family, filling out the torus surface.

As a particular example, the dynamics shown in Fig.~\ref{fig:6}(a-b) are in the form of 
time series from $N=100$ equidistributed spatial points, each of total length $T=500$ (the periods of the two oscillations are $\approx 56$ and $\approx 250$).
These time series are then subdivided into partially overlapping windows of length $l_{\text{string}}=100$, at several degrees of overlap, as will be discussed below.
For an overlap of $l_{\text{string}}-n$ time steps and evenly distributed segments in time, this will give a total of $(T-l_{\text{string}})/n+1$ time windows.

How the torus is reconstructed by applying Diffusion Maps to these time windows, will, as in the previous sections, depend on the scale of the observer: the kernel parameter $\epsilon$.
For $\epsilon=e^{-0.5}\approx 1.7 \cdot 10^{-2}D_{\text{max}}^2$, the dynamics projected on the first three non-trivial Diffusion Map coordinates visually appear one-dimensional: a closed loop spanned by $\phi_1$ and $\phi_2$ (see Fig.~\ref{fig:6}(d)). 
None of the next at least $8$ Diffusion Map coordinates encode new, independent directions.

A single visible ring indicates that, at this kernel scale, only one of the two torus frequencies is 
observable. 
While at this value of $\epsilon$, the temporal modulation is simply missed by the Diffusion Maps to leading order;
the traveling in the spatial domain corresponds to the rotation along this ring.
The last $100$ points plotted correspond to the $100$ time windows obtained from the $100$ different spatial locations \emph{during our last sampling interval in time}.
The smoothness of the coloring by the spatial coordinates at which these observations were made, shows that the Diffusion Maps can recover the physical space $x$.

Decreasing the scale parameter, the ring in Diffusion Map space partially unfolds 
into several rings, as the kernel allows us to begin detecting variability along the modulation direction.
For $\epsilon=e^{-2.0}\approx 3.8 \cdot 10^{-3} D_{\text{max}}^2$, this is shown in Fig.~\ref{fig:6}(e). 
Note that each of the rings is still parametrized by physical space. 
A further decrease of $\epsilon$ leads to further unfolding of the rings, see Figs.~\ref{fig:6}(f-g). 
At $\epsilon=e^{-3.2} \approx 1.2 \cdot 10^{-3} D_{\text{max}}^2$ we get a full separation of the 
rings, making the entire torus finally visible in Fig.~\ref{fig:6}(h).
The direction along each of the rings corresponds to the underlying traveling wave.
We assert that the other direction, from one ring to the next, corresponds to the modulating oscillation:
The time difference between two successive rings is one increment: $n=8$ time units.
Running through seven such overlaps yields a total time length of ${7\cdot8=56}$ time steps, the modulation period.
A finer discretization in time and larger overlaps would clearly ``fill in'' the torus surface.
In the opposite direction, less overlap will ``depopulate'' the torus surface, see Fig.~\ref{fig:6}(i).

A strong relationship exists between our varying the scale of the observer, and topological 
data mining/persistent homology~\cite{kaczynski2004computational}.
Successive plateaus in the identified surface genus in the latter correspond to successive plateaus in 
the dimension of our identified manifold: from a cloud of individual discretization points at very small 
$\epsilon$, to the (desired) torus, then to a ring, and, ultimately, to a single point at very large 
$\epsilon$.
Our Figs.~\ref{fig:6}(d-h) provide an interesting study of the transition between two of the plateaus, 
where the dimensionality of the manifold appears to vary along it~\cite{allarda-2012}. 

In what follows, we will assume that we have chosen the scale at which we want to observe the system, and 
revisit the effect of different types of observations.
The dynamics live on a $T^2$ in function space. We want to be able to observe this $T^2$, that is, we want a mapping between each of our observations and a corresponding unique point on the surface of this torus.
How many quantities (variables) do we need to observe to construct such a mapping?
This has been a long-standing research issue in topology as well as in dynamical systems.
Starting with Whitney's theorem in the 1930s~\cite{whitney}, guaranteeing homeomorphisms between manifolds and their Euclidean embeddings, and through the work of Nash in the 1950s providing isometries~\cite{nash-1954}, we have the work of Takens in 1980~\cite{takens-1981} (and also of Farmer et al.~\cite{Packard}) guaranteeing homeomorphisms between the system state space and a space spanned by just a few delayed measurements of a single scalar observable (see~\cite{sauer-1991}).
This is exactly what we did, using time windows at various spatial locations in our discussion above.
Since the dimension of the attractor is $2$, our $99$ delays were certainly sufficient to guarantee an embedding.

In fact, what we observe is the function $u(x, t)$ for $x \in [0, 2\pi]$ and $t$ within the modulation period. 
This is a two-parameter family of points, and, at each one of them, we took short temporal measurements 
(a value and $99$ delays).
Clearly, there are other ways by which one can learn (observe) this surface. One can take horizontal measurement segments (short discretized solution profiles at fixed time). 
One can consider traveling observers: short measurement segments along an arbitrary angle in $(x,t)$.
Alternatively, as discussed above, we can consider small {\em spatiotemporal patches} centered at each point
(or, to make the observation finite, several leading Fourier components of the function within each patch, 
or several leading PCA components of the histogram of the function values in the patches).
Each of these different types of measurement data simply constitutes a different way to observe ``the same'' points 
on the attractor in function space (or, alternatively, points on our $u(x, t)$ surface).
All of these different types of measurements of points on the attractor (all these different embeddings of the original torus) can therefore be diffeomorphically mapped to each other.
\begin{figure}[!ht]
  \centering
    \includegraphics[width=\columnwidth]{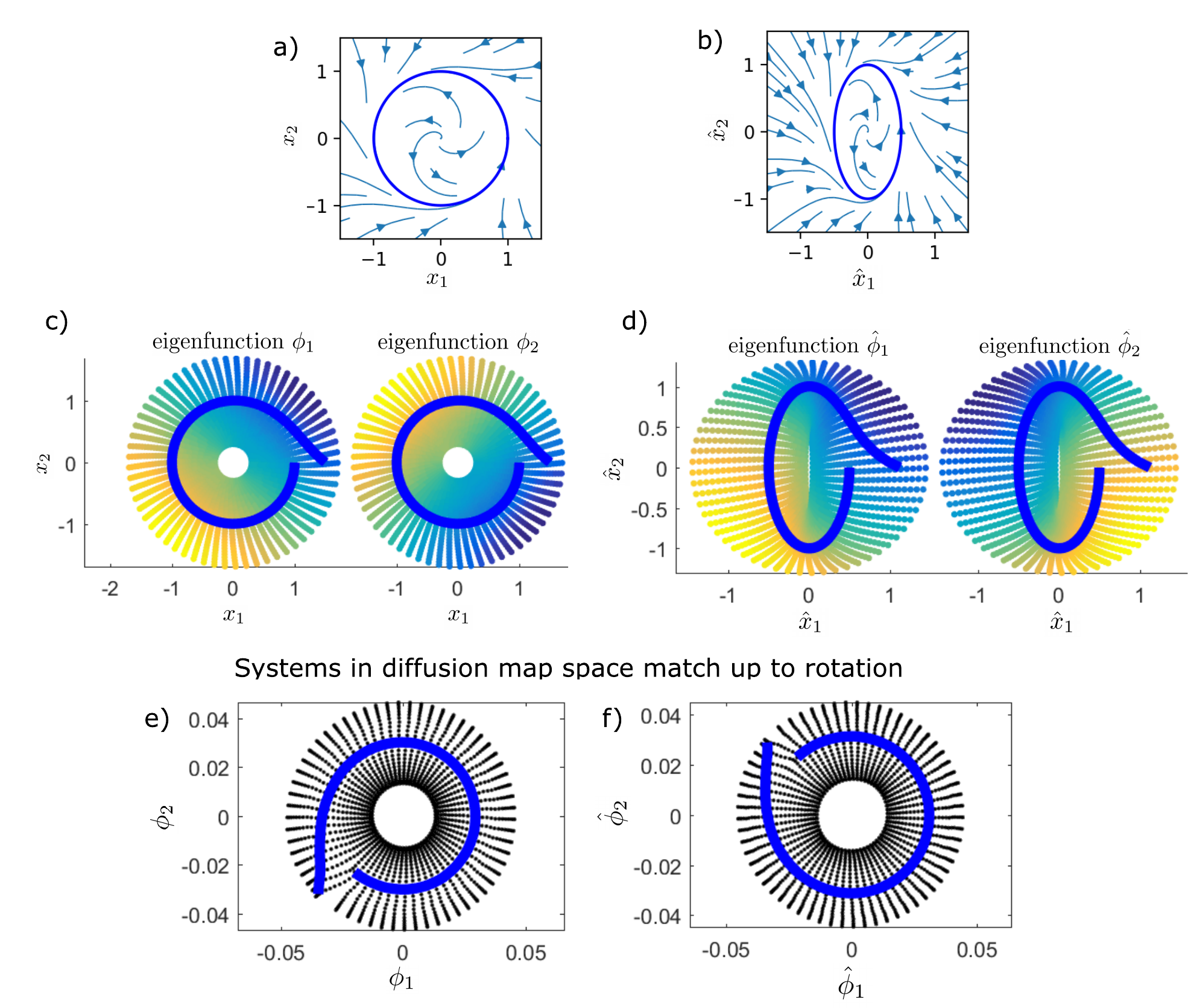}
  \caption{Panel (a) shows a limit cycle system with $\dot{\theta}=1,\dot{r}=r-r^3$, where we add small amplitude white noise at each time step (isotropic in Cartesian space, not visible in the plots). Panel (b) shows the same limit cycle, but observed through $T(r,\theta)=[(r^2)/2\cos(\theta),\sqrt{r}\sin(\theta)]$.
The second row shows the Diffusion Map eigenvectors obtained from the original system, $\phi_{1,2}$, and the transformed system, $\hat{\phi}_{1,2}$, as color on the phase space around the limit cycle. An example trajectory is depicted in blue.  The third row shows the two systems in Diffusion Map space, where they match up to a rotation because we rescaled the distances based on the local covariances of the noise.
    }
  \label{fig:mahalanobis_limitcycle}
\end{figure}

Interestingly, if we can agree on a reference set of observables and construct the conjugacy between this 
set and our various types of observation, we can then easily fuse information from the various 
types of observation (i.e. the different measurement instruments/modalities).
It would make sense for this reference set of observables to be in some sense intrinsic to the manifold,
and not to depend on the particular embedding;
observables based on the manifold curvature tensor, like the ones 
resulting from the Codazzi-Mainardi equations, might be good candidates for this~\cite{Blaschke1973}.

Manifold learning can therefore go beyond the recovery of physical space or the creation of useful embeddings for dynamical observations. It holds the promise of fusing heterogeneous observations of the same dynamical system and of realizing when different dynamical systems are indeed observations of each other (are conjugate).  
The property of obtaining a description of a system that does not depend on the 
measurement instrument is sometimes termed ``gauge invariance'', and indeed performing gauge-invariant 
data mining constitutes a promising research direction.
Diffusion maps based on a Mahalanobis-like distance (involving the local pseudo-inverse of the noise covariance) were proposed by Singer and Coifman in 2008~\cite{SINGER2008226},
and have been used to this effect in e.g. \cite{Singer22092009, talmon13_empir_intrin_geomet_nonlin_model, 
dsilva16_data_driven_reduc_class_multis}.

As a last example, Fig.~\ref{fig:mahalanobis_limitcycle} demonstrates this data-driven matching of a simple stochastic nonlinear oscillator (system (a)) with a nonlinear observation of it (system (b))---a representative example of domain adaptation (i.e., adaptation of a model to a new application domain). 
We parametrize the states $x=(x_1,x_2)$ and $\hat{x}=(\hat{x}_1,\hat{x}_2)=T(r(x_1,x_2),\theta(x_1,x_2))$ of both systems separately with Diffusion Maps. Instead of the Euclidean distance, we use the Mahalanobis distance between data points $x^i,x^k$ (and $\hat{x}^i,\hat{x}^k$, respectively), defined through~\cite{SINGER2008226}
\begin{equation}
d({x}^i,{x}^k)^2=\frac{1}{2}({x}^i-{x}^k)^T(\textbf{C}(x^i)+\textbf{C}(x^k))^\dagger(x^i-x^k).
\end{equation}
Here, $\textbf{C}(x^i)$ and $\textbf{C}(x^i)$ are covariance matrices obtained by short bursts of trajectories started at the given points. In the original system, the short bursts result in isotropic diffusion (so that the covariance is the identity), while the observation function distorts this to anisotropic diffusion in system (b).
The pseudo-inverse $(\cdot)^\dagger$ is used in general, because the matrices are usually rank-deficient. 
The local noise covariance provides an estimate of the local Jacobian of the transformation that is used to re-scale the local distances, such that the metrics on both spaces agree. Parametrization with Diffusion Maps then allows us to match the two dynamical systems modulo a rotation. 
An alternative approach to creating data-driven conjugacies between dynamical systems using Koopman operator 
eigenfunctions~\cite{budisic-2012} is discussed in the Supplemental Information.


\section{Discussion}
We started by demonstrating that manifold learning techniques and, in particular, Diffusion Maps, can be used to reconstruct the topology of the physical space in which an observed process takes place, only from disorganized (spatially unlabelled) collections of temporal measurements (time series segments).
This becomes especially useful in problems where no obvious physical space is involved, such as network dynamics. Here, by discovering intrinsic variability directions in the dynamics, the network can be conveniently reduced: Instead of thousands of coupled oscillators, the response surface can be described in terms of a few collective Diffusion Map coordinates in our Emergent Space.

We then focused on chimera states arising in the dynamics of integrally coupled partial differential equations. 
Being able to tune systematically the kernel parameter $\epsilon$ gives us the ability to vary the scale of the observer and, ultimately, to construct representations of the system at different levels of coarse-graining~\cite{ZelnikManor2004SelfTuningSC,hein05_intrin_rd,Coifman2008}.
In our examples, we always use the same value of $\epsilon$ for all data points of a given data set. Note that for data where the 
density of points varies greatly over the data set, or where the density cannot be bounded from below, 
variable bandwidth kernels can be constructed~\cite{Berry201668}.

We also discussed the observation of the same physical space topology through several different 
types of measurements: short time series windows, short spatial segments, small space-time patches.
The ability to fuse heterogeneous observations of a dynamical system naturally brought forth the possibility of realizing that different dynamical systems are really observations of each other.

The question then arises: Given two dynamical systems, are their state spaces/their representations related by a diffeomorphism, such that the dynamics in the first system correspond to the dynamics in the second?
This question is vital in the theory of dynamical systems, where such diffeomorphisms are typically constructed analytically. 
A famous closed-form example of such a transformation is the Cole-Hopf transform between the diffusion equation and the viscous Burgers equation~\cite{ColeHopf}. 
The related idea of a normal form, shared by two systems, also provides an avenue towards matching them.
We illustrated a data-driven path towards such a matching without access to closed-form models, 
but rather via processing observations of both systems' states with the help of a Mahalanobis-like distance in the Diffusion Map 
computations. 
Any model, even a quite inaccurate one, that lies in the same universality class as the true process we observe 
(that shares the same normal form with the true process), is but one diffeomorphism away from the truth.
Modern machine-learning tools (like deep artificial neural networks) can help find this diffeomorphism in a data-driven way. 
This ``calibration curve'' is a matching between the inaccurate model and the truth. 
Such calibration curves arise in the recent multifidelity modeling literature, e.g.~\cite{perdikaris-2017}.

If machine learning (gauge-invariant data mining) allows us to work practically indiscriminately 
with any possible diffeomorphic realization of a system, which one should we choose?
Which one is ``the best''?
Until now, the optimality criteria were guided by human understanding of the process. 
Choosing physically interpretable variables for a model makes it easier to understand.
It was this anthropocentric sense of simplicity that decided the observables and the form of the model equations. 
However, the temporal evolution in these variables might be rather complicated. 
Lax pairs provide an alternative point of view in choosing one's sense of ``simplicity'': here, 
the variables are quite complicated objects, resulting from eigenproblem solutions at every temporal step, 
yet the dynamics are almost trivial~\cite{lax-1968}.

Machine learning can help us broaden our scope of possible simplicities, in that we can now choose from 
a much wider range of different system representations, and select the kind of simplicity we prefer: 
for example, linear dynamics in a minimal number of variables, if possible.
This is emphatically not a new idea. Principal Component Analysis (PCA) is a time-honored reduction method. 
``Nonlinear'' principal components (the auto-encoder bottleneck neural networks of the 
1980s~\cite{KRAMER1992313} that are now back as deep artificial neural networks) can also provide 
parsimonious descriptions of processes in terms of a few observables (the states of the bottleneck 
neurons, that are, however, not easily physically interpretable).
What has made these techniques widely applicable is the computational savings in producing accurate predictions, despite the ``uninterpretability'' of their state variables.
It is reassuring that ``easy'' back-and-forth mappings between these and the interpretable, 
physical variables can be constructed.

We close this paper with a small gallery of examples, where we chose simplicity to mean a sense of beauty 
(admittedly, our own subjective criterion!). These examples are discussed in more detail in the Supplementary Material. 
The first system is a chimera state, containing, in interpretable variables, spatiotemporal chaos. 
In Emergent Space, it appears to admit periodic, or, at worst, quasi-periodic, visually remarkably 
coherent motion [\href{https://github.com/fkemeth/Equal_Space_Videos/blob/master/vid_1.avi}{Video 1}]].
The second system is also a chimera state; 
in Emergent Space it brings to the minds of each of us certain distinctive swarming behavior of 
the starlings of Rome [\href{https://github.com/fkemeth/Equal_Space_Videos/blob/master/vid_2.avi}{Video 2}].
The last example system consists of 10000 globally coupled Stuart-Landau oscillators. 
In Emergent Space, their probability density function evolves on what visibly suggests a homoclinic 
tangle involving the stable and unstable manifolds of a coarse periodic solution [\href{https://github.com/fkemeth/Equal_Space_Videos/blob/master/vid_3.avi}{Video 3}]. 
In the words of the Little Prince, ``c'est v\'{e}ritablement utile puisque c'est joli''~\cite{de2001petit}.

The work we have discussed has an increasingly medieval flavor: We aspire to predictions, 
starting from observations, processing the data and never trying to grasp the physical nature of the observables 
or the interactions between them (obtaining closed-form expressions for physical laws).

It would appear that because of the complexity and interconnectedness of the systems we study today, the emphasis gradually shifts from understanding \emph{the system and its laws} as a whole, to understanding \emph{the algorithms} that create representations of the system, based on a possibly new, different sense of simplicity. 
Even though the new, data-driven variables may be difficult to rationalize, the mathematics involved in using them to make predictions remain the same.

\section*{Acknoledgements}
The authors thank Robert Axelrod, Ronald Coifman, Maximilian Patzauer, Munir Salman 
and Juan Bello-Rivas for fruitful discussions. 
Financial support by the 
\textit{Institute for Advanced Study - Technische Universit\"{a}t
M\"{u}nchen},
funded by the German Excellence Initiative, by the US National Science Foundation and by DARPA 
is gratefully acknowledged.

\bibliographystyle{unsrt}
\bibliography{lit_corrected}

\end{document}